%% file: workshop_paper.tex
\documentclass{article}
\usepackage{iclr2026_conference,times}
\input{math_commands.tex}

\usepackage{amssymb}
\usepackage{hyperref}
\usepackage{url}
\usepackage{booktabs}
\usepackage{graphicx}

\newtheorem{definition}{Definition}

\title{Toward a Theory of Hierarchical Memory for Language Agents}

\author{
  Yashar Talebirad \\
  University of Alberta \\
  \texttt{talebira@ualberta.ca} \\
  \And
  Ali Parsaee \\
  University of Alberta \\
  \texttt{Parsaee@ualberta.ca} \\
  \And
  Csongor Y. Szepesvari \\
  University of Alberta \\
  \texttt{csongor@ualberta.ca} \\
  \And
  Amirhossein Nadiri \\
  York University \\
  \texttt{anadiri@yorku.ca} \\
  \And
  Osmar Zaiane \\
  University of Alberta \\
  \texttt{zaiane@ualberta.ca}
}
\iclrfinalcopy
\begin{document}
\maketitle

\begin{abstract}
Many recent long-context and agentic systems address context-length limitations by adding \emph{hierarchical memory}: they extract atomic units from raw data, build multi-level representatives by grouping and compression, and traverse this structure to retrieve content under a token budget. Despite recurring implementations, there is no shared formalism for comparing design choices. We propose a unifying theory in terms of three operators. \textbf{Extraction} ($\alpha$) maps raw data to atomic information units; \textbf{coarsening} ($C = (\pi, \rho)$) partitions units and assigns a representative to each group; and \textbf{traversal} ($\tau$) selects which units to include in context given a query and budget. We identify a \textbf{self-sufficiency} spectrum for the representative function $\rho$ and show how it constrains viable retrieval strategies (a \textbf{coarsening-traversal coupling}). Finally, we instantiate the decomposition on eleven existing systems spanning document hierarchies, conversational memory, and agent execution traces, showcasing its generality.
\end{abstract}

\section{Introduction}

Language agents increasingly rely on large external corpora and long interaction histories, but larger context windows do not reliably improve information use. Models underweight mid-context evidence~\citep{liu2024lostmiddle}, struggle with single-fact retrieval in long documents~\citep{hsieh2024ruler}, and often degrade as context grows (``context dilution''/``context rot''). This reveals a gap between \emph{capacity} (more tokens) and \emph{control} (what to surface, and when). Hierarchical memory is a recurring response: RAPTOR~\citep{sarthi2024raptor}, GraphRAG~\citep{edge2024graphrag}, xMemory~\citep{xmemory2025}, H-MEM~\citep{hmem2025}, and SimpleMem~\citep{simplemem2025} build multi-level representations by grouping, compressing, and selectively expanding under a budget; analogous ideas appear in agent execution-trace memory, including reasoning-focused traces (MemoBrain~\citep{memobrain2025}, StackPlanner~\citep{stackplanner2025}) and action/state traces (InfiAgent~\citep{infiagent2025}). Yet these systems are usually presented as isolated design points,  making it hard to compare assumptions.

\begin{figure}[t]
\centering
\includegraphics[width=0.45\textwidth]{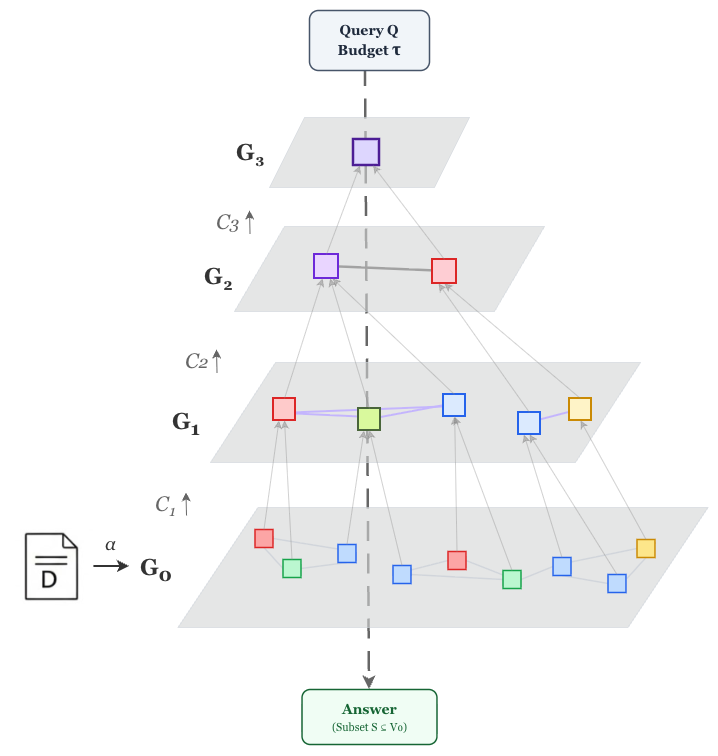}
\caption{The $(\alpha, C, \tau)$ pipeline: $D$ is extracted ($\alpha$) into atoms $G_0$; coarsening $C_1, C_2, C_3$ yields layers $G_1, G_2, G_3$; traversal $\tau$ takes a query and budget and returns a subset $S$ of atoms.}
\label{fig:pipeline}
\end{figure}

We show that, every hierarchical memory system, whether it operates on stored knowledge or live agent execution traces, instantiates a single three-operator pipeline (Figure~\ref{fig:pipeline}). Extraction ($\alpha$) maps raw data to a graph of atomic information units. Coarsening ($C = (\pi, \rho)$) partitions units into groups and produces a representative for each group, yielding a smaller graph, and iterating this produces the hierarchy. Traversal ($\tau$) takes the hierarchy, a query, and token budget, and returns a subset of atomic units to include in context.
The central theoretical observation is that the representative function $\rho$ varies along a self-sufficiency spectrum: a detailed summary preserves most of the group's information, while a category label preserves almost none.
This property constrains which retrieval strategy works. Self-sufficient representatives support collapsed search over all levels (RAPTOR), while referential representatives require top-down refinement (H-MEM, xMemory). We call this constraint the coarsening-traversal (C--T) coupling.

Prior efforts classify memory systems but do not formalize hierarchy construction. CoALA~\citep{sumers2024coala} organizes memory by cognitive type (working, episodic, semantic, procedural), and MemEngine~\citep{memengine2025} implements multiple systems from shared primitives. We provide a mathematical unification, extended to agent execution traces: a formal $(\alpha, C, \tau)$ decomposition with information monotonicity, a self-sufficiency spectrum with C--T coupling, and an instantiation of this decomposition across eleven data and trace systems.

\section{The Framework}

\subsection{Core Definitions}

Natural language exhibits multiscale structure (themes, topics, individual facts), and queries may target any scale. We define three operators that build multiresolution representations of this structure.

\textbf{Information units.} Let $\mathcal{F} = \mathcal{F}_1 \times \cdots \times \mathcal{F}_p$ be a product of feature spaces with one distinguished factor $\mathcal{F}_c = \Sigma^*$ (text). A \emph{unit} $u \in \mathcal{F}$ has projections $\phi_i(u)$; $\phi_c(u)$ is the content. Typical attributes include embeddings $\phi_{\mathrm{emb}}$, timestamps $\phi_{\mathrm{time}}$, entities $\phi_{\mathrm{ent}}$, and structural paths $\phi_{\mathrm{path}}$.

\textbf{Unit graph.} $G = (U, E)$ with $U$ a finite set of units and $E \subseteq U \times U$. The edge set may be empty (a flat bag of units, as in SimpleMem), may encode entity-entity relationships (a knowledge graph, as in GraphRAG), or may encode sequential adjacency (a conversation stream, as in xMemory).

\textbf{Extraction.} The first operator maps raw data into structured units. $\alpha : \mathcal{D} \to \mathcal{G}$ maps raw data $D$ to a unit graph $G_0 = (U_0, E_0)$. All preprocessing (chunking, named entity recognition, path assignment) happens inside $\alpha$, and thus, each unit is self-contained in the resulting graph $G_0$.

\textbf{Coarsening.} A coarsening operator on $G=(U,E)$ is $C=(\pi,\rho)$: (i) $\pi:U \twoheadrightarrow [m]$ is a surjective \emph{grouping} that induces the partition $\{G_j=\pi^{-1}(j)\}$; (ii) $\rho:2^U \to \mathcal{F}$ is a \emph{representative function} such that $\rho(G_j)\in\mathcal{F}$. We assume $|U'|=m<|U|$, where $U'=\{\rho(G_j): j\in[m]\}$.\footnote{Some systems (RAPTOR's GMM soft clustering, overlapping community detection) allow a unit to belong to multiple groups, replacing $\pi:U\to[m]$ with $\pi:U\to 2^{[m]}\setminus\emptyset$; the hierarchy then becomes a DAG.}

\textbf{Hierarchy.} Iterating $C_i$ produces a hierarchy $\mathcal{H} = (G_0, C_1, \ldots, C_L)$ with $G_\ell = C_\ell(G_{\ell-1})$.\footnote{$\mathcal{H}$ lists the initial graph $G_0$ and the coarsening \emph{operators}; the graphs $G_1, \ldots, G_L$ are the \emph{results} of applying them, not separate entries in the tuple. The ``layers'' in the multi-level view are these derived graphs.}

Denote the node set of $G_\ell$ as $V_\ell$. Level $0$ is thus the atom set $V_0$; each application of $C_\ell$ builds the next layer $V_\ell$ from $V_{\ell-1}$ by grouping nodes (via $\pi_\ell$) and assigning one representative per group (via $\rho_\ell$), with edges from each representative to its children. The hierarchy can thus be viewed as a multi-layer graph: the bottom layer holds fine-grained units, and each higher layer is a coarsened graph whose nodes summarize the subgraph below (Figure~\ref{fig:pipeline}). Each $C_\ell$ may use a different grouping and representative strategy (e.g., xMemory uses temporal grouping at level 1, semantic grouping at level 2, and thematic grouping at level 3).

\subsection{Multi-Resolution Representation}
\label{sec:multires}

Assume the input data $D$ is drawn from a distribution over $\mathcal{D}$. Since $V_0 = \alpha(D)$ and each $V_\ell = C_\ell(V_{\ell-1})$, every level inherits randomness from $D$, and the entropies below are well defined. Let $\mathcal{H} = (G_0, C_1, \ldots, C_L)$, and assume each $C_\ell$ is deterministic and information-losing with positive probability under $D$ (equivalently, not almost surely invertible). Then: (i) $H(V_0) > H(V_1) > \cdots > H(V_L)$, since determinism gives $H(V_\ell) \le H(V_{\ell-1})$ and strict information loss on a nonzero-probability set makes the inequality strict; (ii) for any query variable $Q$ with query-agnostic coarsening (i.e., $V_\ell$ depends only on $V_{\ell-1}$ and not on $Q$), $I(Q; V_0) \ge I(Q; V_1) \ge \cdots \ge I(Q; V_L)$, since learning $V_\ell$ provides no additional information about $Q$ once $V_{\ell-1}$ is known; equivalently, $Q$--$V_{\ell-1}$--$V_\ell$ forms a Markov chain (and the Data Processing Inequality applies).

This formalizes that every coarsening step loses information.\footnote{This also aligns with the Information Bottleneck principle~\citep{tishby2000ib}: as compression becomes more aggressive at each level of $\mathcal{H}$, representations move along a trade-off curve between retained detail and compactness, suggesting a multi-resolution structure.} The atoms $V_0$ set the information ceiling; finer extraction increases $H(V_0)$ and raises the ceiling on downstream retrieval quality.\footnote{Appendix~\ref{app:fano} relates data size, context window, and compression ratio to the number of levels needed.}

\subsection{Self-Sufficiency and the C--T Coupling}
\label{sec:ct_coupling}

The representative function $\rho$ is where systems diverge most, and its properties determine the behavior of the entire hierarchy. The \textbf{self-sufficiency} of $\rho$ at group $G_j$ measures how much of the group's information the representative preserves:
\footnote{A computable LLM-based $\mathrm{SS}_\theta$ and a query-dependent refinement $\mathrm{SS}_Q$ are developed in Appendix~\ref{app:ss_theta}.}
\begin{equation}
\mathrm{SS}(\rho, G_j) = \frac{I(G_j; \rho(G_j))}{H(G_j)} = 1 - \frac{H(G_j \mid \rho(G_j))}{H(G_j)}.
\end{equation}
Assume $H(G_j)>0$; if $H(G_j)=0$, define $\mathrm{SS}=1$. When self-sufficiency is high (close to 1), the representative preserves most of the group's content and can answer queries on its own (e.g., LLM-generated summaries). When self-sufficiency is low (close to 0), the representative serves only as a routing label, indicating what the group contains without reproducing it (e.g., domain names). For any query $Q$, $I(Q; G_j) - I(Q; \rho(G_j)) \le H(G_j \mid \rho(G_j))$, so high self-sufficiency bounds the information loss incurred by reading the representative rather than the children.

\paragraph{C--T coupling.} Self-sufficiency constrains which traversal strategies are viable, reflecting a rate-distortion tradeoff. Self-sufficient $\rho$ operates at low distortion: coarse nodes carry enough information to answer queries directly, so collapsed search suffices (single-shot decoding; RAPTOR). Referential $\rho$ operates at high distortion: coarse nodes serve only as routing labels, so top-down refinement is needed to recover the lost detail (successive refinement; H-MEM, xMemory). Mismatching $\rho$ and traversal wastes budget: collapsed search over referential representatives spends tokens on uninformative labels, while top-down routing through self-sufficient representatives spends tokens on unnecessary expansion. Existing evidence is consistent with this observation, but a controlled comparison across the spectrum remains open.

\subsection{Traversal}

A \textbf{traversal} $\tau : (\mathcal{H}, q, B) \to S \subseteq V_0$ takes a hierarchy, query, and token budget and returns atoms satisfying $\sum_{u \in S} |\phi_c(u)| \le B$. By pruning via hierarchy structure, traversal can reduce relevance evaluations from $O(n)$ (flat search) to $O(\log n)$ under bounded compute. Four patterns recur: top-down refinement (H-MEM, xMemory), collapsed search across levels (RAPTOR), multi-view parallel retrieval (SimpleMem), and reasoning-based navigation (PageIndex~\citep{zhang2025pageindex}, MemWalker~\citep{chen2023memwalker}).\footnote{Pseudocode appears in Appendix~\ref{app:algorithms}.} This reflects a dual role of $\mathcal{H}$: with referential $\rho$, it mainly serves as an \emph{index} for top-down routing; with self-sufficient $\rho$, it also serves as a \emph{representation} whose non-leaf nodes often answer broad queries directly (collapsed search).

\section{Instantiation}

\subsection{Data and Trace Systems}

To showcase generality, Table~\ref{tab:systems} maps eleven systems to $(\alpha, C, \tau)$. \textbf{Data memory} systems operate on stored content: batch document hierarchies (RAPTOR, GraphRAG), online conversational memory (xMemory, H-MEM, SimpleMem), observational compression (Mastra's Observational Memory~\citep{barnes2026observational}), and structured navigation (PageIndex). \textbf{Agent execution-trace} systems operate on execution: $\alpha$ segments traces, $C$ groups/compresses them (e.g., MemoBrain's Fold/Flush, StackPlanner's Condensation/Pruning), and $\tau$ reconstructs working context. In trace memory, coherence is causal-functional (steps for the same subproblem) rather than semantic, and the training signal is task reward (DPO, GRPO) rather than retrieval metrics. AgeMem~\citep{agemem2025} exposes store/retrieve/summarize/filter/discard as tool actions, while InfiAgent uses bounded reconstruction from workspace state $\mathcal{F}_t$ and recent actions~\citep{infiagent2025}.\footnote{Bounded context forces a scope--nuance trade-off, which explains convergence on ``capture at full granularity, coarsen iteratively, reconstruct on demand via $\tau$''.}
\begin{table}[t]
\centering
\caption{Data and agent execution-trace systems as $(\alpha, C, \tau)$. Type: D = data, T = trace.}
\label{tab:systems}
\scalebox{0.7}{
\begin{tabular}{@{}llllll@{}}
\toprule
System & Type & $\alpha$ & $\pi$; $\rho$ (SS) & $\tau$ \\
\midrule
RAPTOR & D & 100-tok chunks & UMAP+GMM; LLM summary (high) & collapsed / top-down \\
GraphRAG & D & entity/rel extract & Leiden; community report (high) & global / entity fan-out \\
xMemory & D & episode boundaries & guided split/merge; fact distill (mid) & top-down + expand \\
H-MEM & D & episode + profile & LLM 4-level; domain labels (low) & top-down FAISS \\
SimpleMem & D & window + density & online synthesis; consolidated (high) & multi-view parallel \\
Mastra's OM & D & token threshold & all-in-one; Reflector (high) & full log \\
PageIndex & D & structural parse & doc structure; titles/summaries (mid) & reasoning-based \\
MemoBrain & T & episode $\to$ thought & dep.-subgraph; Fold (high) / Flush (low) & context projection \\
StackPlanner & T & task stack entries & segment; Condensation (high) / Pruning (low) & stack top + retrieval \\
AgeMem & T & turns $\to$ entries & learned; SUMMARY (high) / FILTER (low) & RETRIEVE + FILTER \\
InfiAgent & T & artifacts $\to$ files & state consolidation (mid) & bounded $g(\mathcal{F}_t, a_{t-k:t-1})$ \\
\bottomrule
\end{tabular}
}
\end{table}

\section{Discussion and Future Work}
\label{sec:discussion}

The $(\alpha, C, \tau)$ decomposition provides a common language for comparing systems. Table~\ref{tab:systems} shows that data-memory and agent-trace systems converge on the same pipeline despite different motivations. The C--T coupling suggests that $\rho$ and $\tau$ should be chosen jointly, and the Fano bound (Appendix~\ref{app:fano}) quantifies how representative quality caps branching factor. Current grouping functions $\pi$ mostly assume unweighted, deterministic edges, but real unit graphs are often weighted or uncertain. Local community mining and search methods for weighted (SIWOw~\citep{zafarmand2023siwo}) and uncertain (USIWO~\citep{talebirad2023usiwo}) graphs motivate extending $\pi$ to exploit edge strength/confidence and improve partition coherence (Appendix~\ref{app:coherence}).

As agent execution traces grow, most prior steps become irrelevant to the current subtask, creating a similar context pressure that motivates hierarchical organization of stored knowledge. This convergence suggests that complex agentic tasks involve three coupled hierarchies: the data hierarchy $\mathcal{H}_{\mathrm{data}}$ (stored knowledge), the trace hierarchy $\mathcal{H}_{\mathrm{trace}}$ (reasoning, interaction, and action traces), and the task hierarchy $\mathcal{H}_{\mathrm{task}}$ (problem decomposition). Each node in $\mathcal{H}_{\mathrm{task}}$ induces a traversal through $\mathcal{H}_{\mathrm{data}}$ and generates nodes in $\mathcal{H}_{\mathrm{trace}}$; the inter-hierarchy alignment constraints remain open.

The central limitation is the assumption of a static hierarchy. In practice, hierarchies evolve via insertion/restructuring (xMemory's split/merge, MemTree~\citep{rezazadeh2025memtree}), and retrieval can reshape memory by reinforcement or decay. Both dynamics break the Markov-chain argument of Section~\ref{sec:multires}: query-conditioned $\rho$ makes $V_\ell$ depend on $Q$, and stateful coarsening couples $\tau$ and $C$ into a feedback loop. Formalizing this setting (likely via adaptive information theory or online learning) is the most pressing open problem. Empirical next steps are validating $\mathrm{SS}_\theta$ as a predictor of retrieval quality, testing C--T coupling with matched $\rho/\tau$ variants, and computing IB-optimal $\rho$.
\bibliography{iclr2026_conference}
\bibliographystyle{iclr2026_conference}

\appendix

\section{Alternate Self-Sufficiency Measures}
\label{app:ss_theta}

To make the C--T coupling testable in practice, we need proxies for self-sufficiency that can be estimated on real systems. The measures below separate what can be assessed at construction time from what is revealed only under concrete queries.

\paragraph{Computable proxy ($\mathrm{SS}_\theta$).}
Shannon entropy is not directly computable for natural language, so we use an LLM $P_\theta$ as a practical proxy:
\begin{equation}
\mathrm{SS}_\theta(\rho, G_j) \;=\; 1 - \frac{-\log P_\theta(G_j \mid \rho(G_j))}{-\log P_\theta(G_j)}.
\end{equation}
Intuitively, this measures how much the representative helps the model predict the group's contents. Since many systems use LLMs to generate representatives, $\mathrm{SS}_\theta$ directly affects downstream quality. If $\rho$ hallucinates (adds facts not supported by the children), then $P_\theta(G_j \mid \rho)$ drops: the representative no longer predicts the actual children and may actively mislead. The same drop occurs if $\rho$ blurs important distinctions (e.g., by merging dissimilar items into a vague summary), because the representative fails to disambiguate.

These failure modes suggest that $\mathrm{SS}_\theta$ is a reasonable signal of the reliability of $\rho$. The decision rule linking SS to traversal choice (Section~\ref{sec:ct_coupling}) holds only when $\mathrm{SS}_\theta$ indicates that $\rho$ is reliable. When $\mathrm{SS}_\theta$ is low (due to aggressive compression, hallucination, or incoherent grouping), the system should default to treating $\rho$ as referential, regardless of the nominal compression ratio.

\paragraph{Query-dependent refinement ($\mathrm{SS}_Q$).}
\label{app:ss_q}
Both SS and $\mathrm{SS}_\theta$ are query-independent: they measure the fraction of a group's total information preserved by $\rho$. A query $Q$ may need only a small portion of $G_j$'s information. The \textbf{query-dependent self-sufficiency}
\begin{equation}
\mathrm{SS}_Q(\rho, G_j, Q) \;=\; \frac{I(Q;\, \rho(G_j))}{I(Q;\, G_j)}
\end{equation}
measures the fraction of query-relevant information preserved by $\rho(G_j)$ when $I(Q;G_j)>0$ (if $I(Q;G_j)=0$, define $\mathrm{SS}_Q=1$). Since $\rho(G_j)$ is a deterministic function of $G_j$, the chain $Q$--$G_j$--$\rho(G_j)$ is Markov and the Data Processing Inequality gives $\mathrm{SS}_Q \in [0,1]$.

Define query relevance $r_Q = I(Q; G_j)/H(G_j)$. From $I(Q; G_j \mid \rho(G_j)) \leq H(G_j \mid \rho(G_j))$ and the identity $\mathrm{SS}_Q = 1 - I(Q; G_j \mid \rho(G_j))\, /\, I(Q; G_j)$:
\begin{equation}
\mathrm{SS}_Q \;\geq\; 1 - \frac{1 - \mathrm{SS}}{r_Q}.
\end{equation}
When $r_Q$ is large, SS is a tight lower bound on $\mathrm{SS}_Q$. When $r_Q$ is small, the bound does not provide a meaningful constraint, and low-SS representatives can still achieve high $\mathrm{SS}_Q$. This explains why referential representatives (category labels, hash keys, B-tree boundary keys) are effective for routing: the routing query requires very little of the group's total information ($r_Q \approx 0$), so even $\mathrm{SS} \approx 0$ is compatible with $\mathrm{SS}_Q \approx 1$.

SS is the appropriate construction-time metric, since no query is available when $\rho$ is built. At design time, SS determines the default traversal strategy via the C--T coupling (Section~\ref{sec:ct_coupling}). At query time, $\mathrm{SS}_Q$ determines whether the representative suffices or the children must be expanded. Operationally, SS is a prior over expected query-time behavior and should be calibrated on a target workload by comparing SS against empirical means of $\mathrm{SS}_Q$. Accordingly, the C--T coupling is a practical default: high SS suggests more queries are answerable from the representative (collapsed search), while low SS suggests more queries require expansion (top-down refinement).

\section{Fano Bound and Branching Factor}
\label{app:fano}

The Fano inequality constrains how many groups a level can contain given the discriminative quality of the representative.

\textbf{Theorem (Fano).} Let $Z \in \{1, \ldots, n_k\}$ be the correct group index and $O$ the observable routing evidence. For any estimator $\hat{Z} = g(O)$ with error probability $p_e = P[\hat{Z} \neq Z]$:
\[
p_e \ge 1 - \frac{I(Z;O) + 1}{\log_2 n_k}.
\]

\textbf{Corollary} (from xMemory~\citep{xmemory2025}). If $I(Z;O) \le B$ bits and $p_e \le \varepsilon$, then $n_k \le 2^{(B+1)/(1-\varepsilon)}$.

In our framework, $B$ is determined by the information content of $\rho$: low-SS representatives (small $B$) force coarser partitions, while high-SS representatives permit finer ones. Composing across $L$ levels with per-level error $\varepsilon_\ell$, the total routing failure probability is bounded by $\sum_\ell \varepsilon_\ell$ (union bound).

\paragraph{Optimal branching.} Assuming a balanced tree, uniform per-node cost, and no caching, top-down retrieval with beam width $k$ and branching factor $b$ has total cost $L \cdot k \cdot b$. With $n=|V_0| = b^L$, substituting $L = \ln n / \ln b$ yields cost $\propto b / \ln b$, which is minimized at $b = e$. The Fano constraint caps $b$ at $2^{(B+1)/(1-\varepsilon)}$; if this is below $e$, the cap binds. The optimal depth is $L^* = \left\lceil \ln n / \ln b^* \right\rceil$.

\paragraph{Depth bound.} A related structural constraint governs hierarchy depth. If total data comprises $N$ tokens, the context window holds $C$ tokens, and each coarsening level compresses content by a factor $r$, the top level fits in context only when $L \ge \lceil \log_r(N/C) \rceil$.

\section{Affinity and Coherence}
\label{app:coherence}
Traversal quality depends not only on how informative representatives are, but also on whether grouping preserves meaningful neighborhood structure. Here, we make the hidden assumption behind top-down pruning more explicit: grouping must preserve local relevance geometry well enough that parent-level decisions remain predictive at child level.
The C--T coupling (Section~\ref{sec:ct_coupling}) concerns the representative $\rho$; the \emph{grouping} $\pi$ also constrains traversal. Every hierarchical system implicitly assumes that units placed in the same group are in some sense alike, whether by embedding similarity (RAPTOR), graph connectivity (GraphRAG), or structural path (PageIndex). Without a principled notion of ``which units belong together,'' the partition is arbitrary and top-down pruning has no guarantee: a low-relevance parent might have highly relevant children in another topic that happened to be grouped with it. We formalize the notion of ``alike'' as an affinity and require that the partition respect it; that is what makes top-down traversal safe. The definitions below are an attempt to make this notion more precise.

\begin{definition}[Affinity]
An affinity on a unit graph $G = (U, E)$ is a symmetric function $W : U \times U \to \mathbb{R}_{\ge 0}$.
\end{definition}

\begin{definition}[$W$-coherent partition]
A partition $\pi : U \twoheadrightarrow [m]$ is $W$-coherent if within-group affinity exceeds between-group affinity:
\[
\mathbb{E}_{u,v : \pi(u) = \pi(v)}[W(u,v)] > \mathbb{E}_{u,v : \pi(u) \neq \pi(v)}[W(u,v)].
\]
\end{definition}

Systems derive $W$ from different signals:

\begin{table}[h]
\centering
\small
\begin{tabular}{@{}lll@{}}
\toprule
System & Affinity $W(u,v)$ & Partition method \\
\midrule
RAPTOR & $\cos(\phi_{\mathrm{emb}}(u), \phi_{\mathrm{emb}}(v))$ & UMAP + GMM \\
GraphRAG & connectivity in $E$ & Leiden community detection \\
H-MEM & LLM-judged domain co-membership & LLM 4-level classification \\
xMemory & cosine + sparsity-semantic score & guided split/merge \\
SimpleMem & semantic + temporal proximity & online LLM synthesis \\
Mastra's OM & temporal contiguity & token-count threshold \\
PageIndex & $|\mathrm{lcp}(\phi_{\mathrm{path}}(u), \phi_{\mathrm{path}}(v))|$ & structural parsing \\
\bottomrule
\end{tabular}
\end{table}

In Table~\ref{tab:systems}, Leiden community mining (used in GraphRAG) is the clearest example of global graph partitioning. The graph-mining literature also studies \emph{community search}, which identifies a local dense subgraph around given query nodes without processing the full graph~\citep{zafarmand2023siwo}. This suggests a local variant of $\pi$: rather than committing to a single global partition at construction time, the system could perform ad-hoc, on-demand coarsening around the regions of the unit graph most relevant to the current query. This way, the quality of $\pi$ depends directly on the fidelity of the graph signals used by the mining step. Furthermore, real-world unit graphs frequently carry edge weights or probabilistic edges encoding uncertainty in the underlying facts, as commonly arises in knowledge-graph predicates with uncertain truth values. Incorporating such richer signals into $W$ can tighten the coherence gap $\gamma_\ell = \mathbb{E}[W_{\mathrm{within}}] - \mathbb{E}[W_{\mathrm{between}}]$ and improve the quality of $\pi$.

The affinity $W$ is why $\pi$ should group certain units together, but $\pi$ need not compute $W$ explicitly. In structural parsing, $\pi$ reads off path prefixes directly. For agent execution traces, a symmetric pairwise affinity is often a poor fit: MemoBrain groups reasoning thoughts by causal co-resolution (thoughts connected via directed dependencies that jointly resolve a subproblem), and StackPlanner groups stack entries by contiguous stack membership. In both cases, the grouping criterion is structural and directed rather than similarity-based, so we omit these systems from the table. The underlying principle still holds, however: $W$-coherence is what makes top-down traversal efficient, and its analogue in trace systems is that steps addressing the same subproblem should cluster, ensuring that parent relevance predicts child relevance.
The key intuition is \emph{relevance monotonicity}: if a parent node has low relevance to the query, its children are unlikely to have high relevance, with the gap controlled by the coherence gap $\gamma_\ell$ at that level. This justifies top-down pruning in H-MEM, xMemory, and MemWalker, and the same logic applies to MemoBrain's Fold and Flush: if a summarized sub-trajectory is irrelevant to the current reasoning state, its constituent steps are also likely irrelevant. Systems with poor coherence (e.g., noisy LLM partitions) should use wider beams or prefer collapsed search. A formal characterization of how $\gamma_\ell$ bounds child relevance conditioned on parent relevance remains open.

\section{Traversal Algorithms}
\label{app:algorithms}

The algorithms below operationalize the same design trade-off discussed in the main text: spend budget on broad routing first or on direct content evidence.

\paragraph{Algorithm 1: Top-down refinement} (H-MEM, xMemory).
Given $\mathcal{H}$, query $q$, budget $B$, beam widths $k_L, \ldots, k_0$: set $S_L \gets \mathrm{Top\text{-}}k_L(V_L, r(q, \cdot))$. For $\ell = L{-}1, \ldots, 0$: expand $\mathrm{candidates} \gets \bigcup_{v \in S_{\ell+1}} \mathrm{children}(v)$, then $S_\ell \gets \mathrm{Top\text{-}}k_\ell(\mathrm{candidates}, r(q, \cdot))$. Return $S_0$ truncated to budget $B$.

\paragraph{Algorithm 2: Collapsed search} (RAPTOR).
Pool all nodes across all levels: $\mathrm{pool} \gets \bigcup_{\ell=0}^{L} V_\ell$. Retrieve $S \gets \mathrm{Top\text{-}}k(\mathrm{pool}, r(q, \cdot))$. Expand any non-leaf in $S$ to its atoms. Return atoms within budget $B$.
\paragraph{Algorithm 3: Reasoning-based navigation} (PageIndex, MemWalker).
Initialize $\mathrm{context} \gets \phi_c(\mathrm{root}(\mathcal{H}))$. While budget is not exhausted: the LLM selects a child node from context given $q$. If the node is a leaf, collect it. Otherwise, expand: $\mathrm{context} \gets \mathrm{context} \cup \phi_c(\mathrm{children}(\mathrm{node}))$. Return collected atoms.
\paragraph{Algorithm 4: Multi-view parallel retrieval}\label{app:alg:multiview} (SimpleMem).
Decompose the query: $(q_{\mathrm{sem}}, q_{\mathrm{lex}}, q_{\mathrm{sym}}, d) \gets \mathrm{Plan}(q)$. Set candidate limit $n \gets f(d)$. Retrieve in parallel: $R_{\mathrm{sem}} \gets \mathrm{Top\text{-}}n(V_0, r_{\mathrm{sem}}(q_{\mathrm{sem}}, \cdot))$, $R_{\mathrm{lex}} \gets \mathrm{Top\text{-}}n(V_0, r_{\mathrm{lex}}(q_{\mathrm{lex}}, \cdot))$, $R_{\mathrm{sym}} \gets \mathrm{Top\text{-}}n(V_0, r_{\mathrm{sym}}(q_{\mathrm{sym}}, \cdot))$. Return $R_{\mathrm{sem}} \cup R_{\mathrm{lex}} \cup R_{\mathrm{sym}}$ truncated to budget $B$.

Algorithm~4 uses a flat multi-view design: each view indexes the same atoms $V_0$ using a different signal (semantic similarity, lexical overlap, symbolic metadata), and taking the union ensures that a relevant atom missed by one view can still be recovered by another. More generally, a system can maintain multiple hierarchies $\{\mathcal{H}_i\}$ over the same atom set $V_0$, each organizing atoms according to a different principle (e.g., one by temporal--thematic coherence, another by embedding similarity). This can be beneficial because any single grouping $\pi$ is inevitably lossy: semantically related atoms may land in different branches, and the severity grows with partition granularity. A second hierarchy provides an alternative traversal path: an atom mis-partitioned in $\mathcal{H}_1$ may be correctly grouped in $\mathcal{H}_2$, so retrieval failure requires mis-partition in both views rather than just one. Multi-attribute queries benefit further because constraints of different types (e.g., topical relevance vs.\ temporal range) decompose naturally into per-hierarchy retrievals, which can be combined via set intersection for precision or union for recall; when constraints are independent, each retrieval can be parallelized.

\section{Task and Agent Hierarchies}
\label{app:task_agent_extension}

In Section~\ref{sec:discussion}, we mention three coupled hierarchies and note that $\mathcal{H}_{\mathrm{task}}$ has dynamics different from data memory. A useful extension is to separate two orthogonal structures: the \emph{task hierarchy} (what to solve) and the \emph{agent hierarchy} (which solver tier handles each part).

Define $\mathcal{G}_{\mathrm{task}} = (V_{\mathrm{task}}, E_{\mathrm{dep}})$ as a directed acyclic graph, which is gradually constructed during execution to represent dependencies among tasks; its hierarchical, coarse-to-fine ordering is denoted as $\mathcal{H}_{\mathrm{task}}$. Let $(A, \le_A)$ represent a set of agent tiers, partially ordered by capability (and usually by cost as well). In this setup, $\mathcal{G}_{\mathrm{task}}$ expresses the structure of the problem by specifying which subtasks depend on others, while $(A, \le_A)$ characterizes the solvers—showing which agent tiers are equipped to handle tasks of increasing difficulty and generality.

Assignment of tasks to agents can then be handled by a routing map $R : V_{\mathrm{task}} \to A$. To ensure coherence, a natural constraint is monotonicity: if $t$ is an ancestor of $t'$ in $\mathcal{G}_{\mathrm{task}}$, then $R(t) \ge_A R(t')$. This means that higher-level planning is given to more capable (typically more expensive) agents, while specific, lower-level execution is delegated to less capable tiers.

\end{document}

%% file: math_commands.tex
\usepackage{amsmath,amsfonts,bm}

\def\eqref#1{equation~\ref{#1}}

\def\1{\bm{1}}

\DeclareMathAlphabet{\mathsfit}{\encodingdefault}{\sfdefault}{m}{sl}
\SetMathAlphabet{\mathsfit}{bold}{\encodingdefault}{\sfdefault}{bx}{n}



%% file: iclr2026_conference.bib
@inproceedings{sarthi2024raptor,
  title = {{RAPTOR}: Recursive Abstractive Processing for Tree-Organized Retrieval},
  author = {Sarthi, Parth and Abdullah, Salman and Tuli, Aditi and Khanna, Shubh and Goldie, Anna and Manning, Christopher D.},
  booktitle = {International Conference on Learning Representations},
  year = {2024}
}

@article{xmemory2025,
  title = {Beyond {RAG} for Agent Memory: Retrieval by Decoupling and Aggregation},
  author = {Hu, Zhanghao and Zhu, Qinglin and Yan, Hanqi and He, Yulan and Gui, Lin},
  journal = {arXiv preprint arXiv:2602.02007},
  year = {2026}
}

@article{hmem2025,
  title = {Hierarchical Memory for High-Efficiency Long-Term Reasoning in {LLM} Agents},
  author = {Sun, Haoran and Zeng, Shaoning},
  journal = {arXiv preprint arXiv:2507.22925},
  year = {2025}
}

@article{simplemem2025,
  title = {{SimpleMem}: Efficient Lifelong Memory for {LLM} Agents},
  author = {Liu, Jiaqi and Su, Yaofeng and Xia, Peng and Han, Siwei and Zheng, Zeyu and Xie, Cihang and Ding, Mingyu and Yao, Huaxiu},
  journal = {arXiv preprint arXiv:2601.02553},
  year = {2026}
}

@article{edge2024graphrag,
  title = {From Local to Global: A Graph {RAG} Approach to Query-Focused Summarization},
  author = {Edge, Darren and Trinh, Ha and Cheng, Newman and Bradley, Joshua and Chao, Alex and Mody, Apurva and Truitt, Steven and Larson, Jonathan},
  journal = {arXiv preprint arXiv:2404.16130},
  year = {2024}
}

@article{liu2024lostmiddle,
  title = {Lost in the Middle: How Language Models Use Long Contexts},
  author = {Liu, Nelson F. and Lin, Kevin and Hewitt, John and Paranjape, Ashwin and Bevilacqua, Michele and Petroni, Fabio and Liang, Percy},
  journal = {Transactions of the Association for Computational Linguistics},
  volume = {12},
  pages = {157--173},
  year = {2024},
  note = {arXiv:2307.03172}
}

@inproceedings{hsieh2024ruler,
  title = {{RULER}: What's the Real Context Size of Your Long-Context Language Models?},
  author = {Hsieh, Cheng-Ping and Sun, Simeng and Kriman, Samuel and Acharya, Shantanu and Rekesh, Dima and Jia, Fei and Zhang, Yang and Ginsburg, Boris},
  booktitle = {Conference on Language Modeling (COLM)},
  year = {2024},
  note = {arXiv:2404.06654}
}

@article{memobrain2025,
  title = {MemoBrain: Executive Memory as an Agentic Brain for Reasoning},
  author = {Qian, Hongjin and Cao, Zhao and Liu, Zheng},
  journal = {arXiv preprint arXiv:2601.08079},
  year = {2026}
}

@article{stackplanner2025,
  title = {{StackPlanner}: A Centralized Hierarchical Multi-Agent System with Task-Experience Memory Management},
  author = {Zhang, Ruizhe and Jiang, Xinke and Yang, Zhibang and Zhang, Zhixin and Gao, Jiaran and Xiao, Yuzhen and Lai, Hongbin and Chu, Xu and Zhao, Junfeng and Wang, Yasha},
  journal = {arXiv preprint arXiv:2601.05890},
  year = {2026}
}

@article{agemem2025,
  title = {Agentic Memory: Learning Unified Long-Term and Short-Term Memory Management for Large Language Model Agents},
  author = {Yu, Yi and Yao, Liuyi and Xie, Yuexiang and Tan, Qingquan and Feng, Jiaqi and Li, Yaliang and Wu, Libing},
  journal = {arXiv preprint arXiv:2601.01885},
  year = {2026},
  note = {AgeMem}
}

@article{infiagent2025,
  title = {InfiAgent: An Infinite-Horizon Framework for General-Purpose Autonomous Agents},
  author = {Yu, Chenglin and Wang, Yuchen and Wang, Songmiao and Yang, Hongxia and Li, Ming},
  journal = {arXiv preprint arXiv:2601.03204},
  year = {2026}
}

@inproceedings{tishby2000ib,
  title = {The Information Bottleneck Method},
  author = {Tishby, Naftali and Pereira, Fernando C. and Bialek, William},
  booktitle = {Proceedings of the 37th Annual Allerton Conference on Communication, Control, and Computing},
  year = {1999}
}

@article{sumers2024coala,
  title = {Cognitive Architectures for Language Agents},
  author = {Sumers, Theodore R. and Yao, Shunyu and Narasimhan, Karthik and Griffiths, Thomas L.},
  journal = {Transactions on Machine Learning Research},
  year = {2024},
  note = {arXiv:2309.02427}
}

@article{memengine2025,
  title = {MemEngine: A Unified and Modular Library for Developing Advanced Memory of {LLM}-based Agents},
  author = {Zhang, Zeyu and Dai, Quanyu and Chen, Xu and Li, Rui and Li, Zhongyang and Dong, Zhenhua},
  journal = {arXiv preprint arXiv:2505.02099},
  year = {2025}
}

@article{chen2023memwalker,
  title = {Walking Down the Memory Maze: Beyond Context Limit through Interactive Reading},
  author = {Chen, Howard and Pasunuru, Ramakanth and Weston, Jason and Celikyilmaz, Asli},
  journal = {arXiv preprint arXiv:2310.05029},
  year = {2023},
  note = {MemWalker; Meta AI Research}
}

@inproceedings{rezazadeh2025memtree,
  title = {From Isolated Conversations to Hierarchical Schemas: Dynamic Tree Memory Representation for {LLMs}},
  author = {Rezazadeh, Alireza and Li, Zichao and Wei, Wei and Bao, Yujia},
  booktitle = {International Conference on Learning Representations},
  year = {2025},
  note = {MemTree; arXiv:2410.14052}
}

@article{zhang2025pageindex,
  author = {Mingtian Zhang and Yu Tang and PageIndex Team},
  title = {PageIndex: Next-Generation Vectorless, Reasoning-based RAG},
  journal = {PageIndex Blog},
  year = {2025},
  month = {September},
  note = {https://pageindex.ai/blog/pageindex-intro},
}

@misc{barnes2026observational,
  author = {Barnes, Tyler},
  title = {Observational Memory: 95\% on LongMemEval},
  howpublished = {Mastra Research Blog},
  year = {2026},
  month = {February},
  url = {https://mastra.ai/research/observational-memory},
  note = {Observational Memory (OM)}
}

@article{zafarmand2023siwo,
  title = {Fast Local Community Discovery Relying on the Strength of Links},
  author = {Zafarmand, Mohammadmahdi and Talebirad, Yashar and Austin, Eric and Largeron, Christine and Za{\"i}ane, Osmar R.},
  journal = {Social Network Analysis and Mining},
  volume = {13},
  number = {1},
  pages = {112},
  year = {2023},
  publisher = {Springer}
}

@inproceedings{talebirad2023usiwo,
  title = {{USIWO}: A Local Community Search Algorithm for Uncertain Graphs},
  author = {Talebirad, Yashar and Zafarmand, Mohammadmahdi and Zaiane, Osmar R. and Largeron, Christine},
  booktitle = {Proceedings of the International Conference on Advances in Social Networks Analysis and Mining},
  pages = {187--194},
  year = {2023}
}
